# Twisted van der Waals Josephson junction based on high-$T_c$ superconductor

Jongyun Lee[1], Wonjun Lee[1], Gi-Yeop Kim[2,3], Yong-Bin Choi[1], Jinho Park[1], Seong Jang[1], Genda Gu[4], Si-Young Choi[2,3], Gil Young Cho[1,5,6], and Gil-Ho Lee[1,6,*], Hu-Jong Lee[1,*]

[1] Department of Physics, Pohang University of Science and Technology, Pohang 37673, Korea

[2] Department of Materials Science and Engineering, Pohang University of Science and Technology, Pohang 37673, Korea.

[3] Materials Imaging & Analysis Center, Pohang University of Science and Technology, Pohang 37673 Korea

[4] Condensed Matter Physics and Materials Science Department, Brookhaven National Laboratory, Upton, New York 11973, USA

[5] Center for Artificial Low Dimensional Electronic Systems, Institute for Basic Science (IBS), Pohang 37673, Korea

[6] Asia Pacific Center for Theoretical Physics, Pohang 37673, Korea

**Stacking two-dimensional van der Waals (vdW) materials rotated with respect to each other show versatility for the study of exotic quantum phenomena. Especially, anisotropic layered materials have great potential for such twistronics applications, providing high tunability. Here, we report anisotropic superconducting order parameters in twisted $Bi_2Sr_2CaCu_2O_{8+x}$ (Bi-2212) vdW junctions with an atomically clean vdW interface, achieved using the microcleave-and-stack technique. The vdW Josephson junctions with twist angles of 0° and 90° showed the maximum Josephson coupling, which was comparable to that of intrinsic Josephson (IJ) junctions in the bulk crystal. As the twist angle approaches 45°, Josephson coupling is**

**suppressed, and eventually disappears at 45°. The observed twist angle dependence of the Josephson coupling can be explained quantitatively by theoretical calculation with the *d*-wave superconducting order parameter of Bi-2212 and finite tunneling incoherence of the junction. Our results revealed the anisotropic nature of Bi-2212 and provided a novel fabrication technique for vdW-based twistronics platforms compatible with air-sensitive vdW materials.**

Van der Waals (vdW) materials provide a variety of electronic phases in two-dimensional (2D) systems, including metals[1-3], superconductors[4-7], insulators[8], semiconductors[9-11], and topological materials[12, 13]. It has become possible to design and engineer material properties by stacking 2D vdW heterostructures. Recently, stacking of vdW materials with a twist angle has provided an opportunity to study a variety of exotic physical phenomena, providing an exciting platform for "twistronics." One good example is twisted bilayer graphene, which shows correlated physics leading to Mott insulator phase, superconductivity, and ferromagnetism[14-17].

$Bi_2Sr_2CaCu_2O_{8+x}$ (Bi-2212) is a mechanically cleavable vdW high-critical temperature superconductor (HTSC)[18, 19]. Bulk Bi-2212 has a layered structure consisting of alternately stacked superconducting $CuO_2$ bilayers and insulating SrO-BiO layers (Figure 1a). The insulating SrO-BiO layers act as a tunnel barrier, forming tunnel-type Josephson coupling between two superconducting $CuO_2$ bilayers along the *c*-axis of the crystal; this is referred to as IJ coupling[20-22]. Such tunnel barrier was used for intrinsic tunneling spectroscopy to examine the symmetry of the superconducting pairing gap of Bi-2212[23, 24]. A recent study using the mechanical cleaving technique demonstrated the 2D nature of cuprate HTSC, even in a monolayer of Bi-2212 with a superconducting transition temperature comparable to the bulk value[7]. A number of studies using angle-resolved photoemission spectroscopy[25-28] and scanning tunneling microscopy[29-32] have suggested the *d*-wave superconducting gap of Bi-2212. In transport measurements, the tunneling Josephson effect between Bi-2212 crystals with well-defined twist angles has been suggested to explain the angular dependence of the superconducting gap[33-35]. Unlike other superconducting materials, Bi-2212 can form a tunneling junction along the *c*-axis without insertion of any insulating layer, because the terminating layer is an insulating BiO layer as mechanical cleavage usually occurs between BiO double layers[36-40] due to the weaker vdW bonding[41, 42]. Therefore, in the last few decades, attempts have been made to experimentally observe the *d*-wave superconducting gap of Bi-2212 using twisted Josephson junctions. However, these studies have yielded inconsistent results regarding the twist angle dependence of Josephson coupling and the superconducting gap. In one study, the twisted junction was formed by repeated mechanical pressing, and the twist angle dependence of the superconducting gap energy showed no significant oscillation[43]. Repeated pressing against two Bi-2212 crystals may cause fractures at the junction, and the surfaces of Bi-2212 that formed the junction may have been degraded in the ambient atmosphere. In another study, twist angle junctions were prepared by sintering slightly below the melting temperature of the bulk crystal for 30 hours, which showed the isotropic twist angle dependence of the critical current density[44]. The junction interface may have been reconstructed near the melting temperature. In contrast, for a junction prepared by stacking two Bi-2212 whiskers and annealing at the whisker growth temperature (850°C)[45], strong angular dependence of the critical current density was observed, but without quantitative theoretical explanation. Recently, the vdW dry-transfer technique has been adopted for studying various Josephson junctions[46-50]. One study reported that the Josephson coupling strength between Bi-2212 crystals appeared to be insensitive to the twist angle[51]. However, the junction interface may have been affected by contamination introduced during the dry-transfer process involving polymers. The limitations of previous studies could be attributed to structural and chemical disorder at the junction interface. Therefore, the fabrication of an atomically clean and sharp junction interface is crucial for investigating the angular dependence of superconducting order in Bi-2212.

In the present study, we fabricated Josephson junctions consisting of two Bi-2212 crystals with controlled twist angle and investigated the anisotropic nature of the superconducting gap of Bi-2212. Using the microcleave-and-stack technique in an inert atmosphere, we could cleave a single crystal of Bi-2212 into two, and then stack them upon each other with a preset twist angle while excluding contaminants from the junction interface. The 0°-twisted Josephson junctions showed Josephson coupling strength comparable to the IJ coupling of the bulk Bi-2212. However, at a twist angle of 45°, the Josephson coupling disappeared within the limits of our experimental resolution, which was expected for the *d*-wave superconducting gap of Bi-2212. The measured twist angle dependence of the Josephson coupling strength was well explained by the theoretical fitting, considering the *d*-wave superconducting gap and incoherent tunneling at the junction.

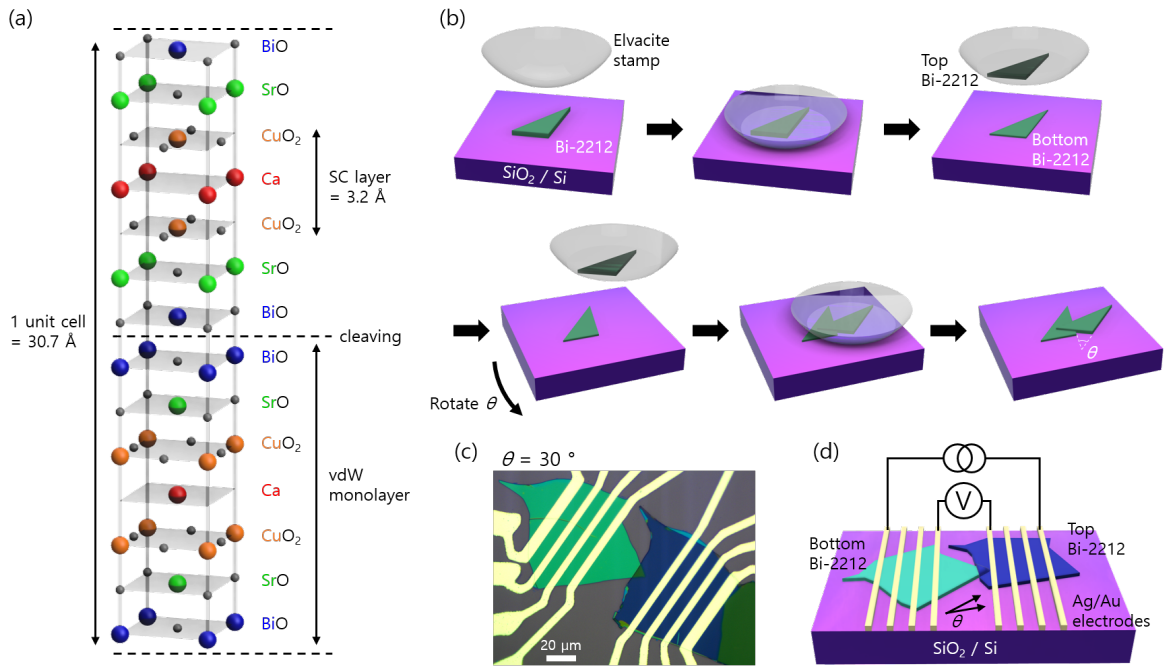

**Figure 1. Fabrication of a $Bi_2Sr_2CaCu_2O_{8+x}$ (Bi-2212) twisted Josephson junction.** (a) Atomic structure of 1 unit cell of Bi-2212. (b) Schematic of the microcleave-and-stack technique for van der Waals (vdW) Josephson junctions. (c) Optical image of a 30°-twisted Josephson junction with Ag/Au electrodes. (d) Schematic and current-biased measurement configuration of the twisted Josephson junction.

To minimize the structural and chemical defects at the vdW interface, a twisted Josephson junction was formed by the microcleave-and-stack technique in a glovebox filled with argon gas (Figure 1b). Optimally doped Bi-2212 single crystals were grown by the traveling floating zone method[52]. First, Bi-2212 flakes on $SiO_2$/Si substrate were prepared by mechanical cleaving using adhesive tape. The Bi-2212 flake was then cleaved again into two pieces using a polymer (thermoplastic methacrylate copolymer; Elvacite 2552C; Lucite International)-based adhesive stamp[53, 54] at 100°C–120°C, in a process we call "microcleaving". The bottom flake was kept on the substrate, while the top flake was attached to the polymer stamp. After the bottom flake had been rotated through the desired angle with respect to the top flake, the top flake was released onto the bottom flake. This process from microcleaving to stacking usually took less than 10 minutes. After releasing the stack of twisted Josephson junctions by melting the stamp at 200°C, polymer residue on the top surface of the stack was removed with acetone, and annealing was performed at 350°C for 30 min in an oxygen atmosphere. The twist angle between the flakes was reexamined under an optical microscope. We emphasize that the bottom surface of the top flake and top surface of the bottom flake did not come into contact with any substances before stacking. We patterned the electrodes by electron-beam lithography, with double-layer resist consisting of 950 K polymethyl methacrylate (PMMA) C4 and copolymer (MMA (8.5) MAA) EL12. The double-layer resist was dried at room temperature without baking. Electrical contact to Bi-2212 flakes was achieved by evaporating Ag/Au electrodes and diffusing Ag into Bi-2212 by annealing at 350°C for several hours[55]. Annealing was performed in an oxygen atmosphere to minimize the loss of oxygen from the Bi-2212 crystal during the process, and Au was overlaid to prevent the oxidation of Ag. This method allowed clean electrical lead contacts to Bi-2212 flakes with areal contact resistance < 300 kΩ·μm$^2$, which was sufficiently low to perform the four-probe measurement. Figure 1c shows an optical microscope image of the finalized device. Figure 1d depicts the schematic of a device and the measurement configuration. A four-probe configuration was used to exclude the contact resistance, and each measurement electrical line was electrically filtered with room-temperature π and RC low-pass filters and cryogenic RC filters, to suppress high-frequency noise.

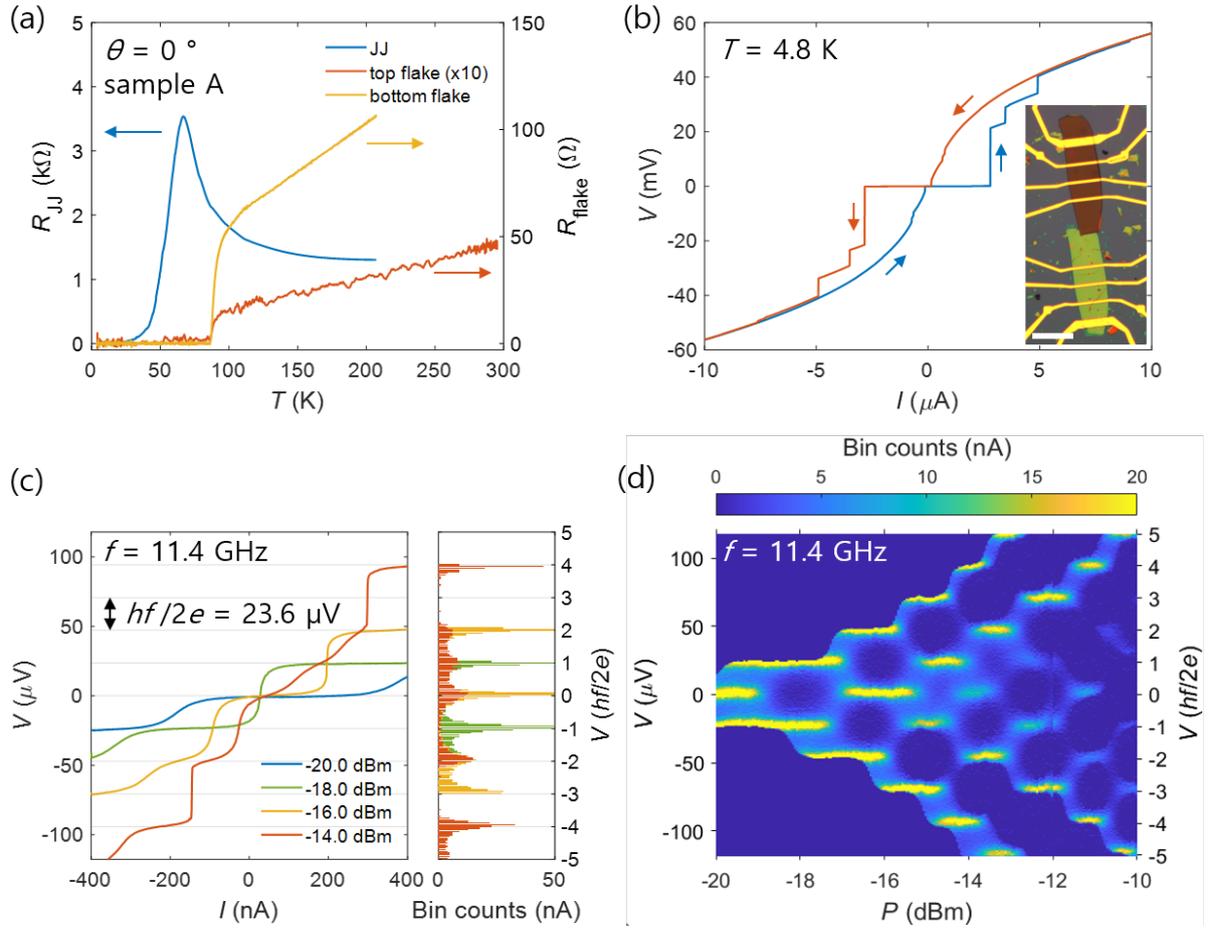

**Figure 2. Basic transport characteristics of a twisted Josephson junction.** (a) Temperature-dependent resistance of the top and bottom flakes and the junction. (b) Tunneling current-voltage (*I-V*) characteristic curve of the junction at temperature *T* = 4.8 K. Current bias was swept from negative to positive current (positive to negative current), shown in blue (orange). The inset is an optical image of the device, from which all data of Figure 2 were measured (scale bar: 50 μm). (c) (Left panel) Shapiro steps in *I-V* characteristic curves under four different microwave powers with fixed frequency *f* = 11.4 GHz. (Right panel) Histogram obtained by binning with a voltage interval of 1.2 μV, for *f* = 11.4 GHz and *P* = −18.0, −16.0, −14.0 dBm. (d) Bin counts map of the junction voltage and the microwave power for frequency *f* = 11.4 GHz.

Figure 2 shows the transport measurement results from a 0°-twisted Josephson junction, indicating representative characteristics of the junctions with different twist angles (for other devices, see Figure S1 in the Supporting Information). Both top and bottom flakes showed superconducting transition at the critical temperature of ~86 K, which was similar to the bulk value (Figure 2a). The temperature dependence of the junction resistance showed insulating behavior above the critical temperature. In addition, the current-voltage (*I-V*) characteristics measured at 4.8 K (Figure 2b) showed suppressed subgap conductance. Such insulating behavior can be explained by tunneling transport across the insulating barriers in the twisted Josephson junction, as well as in the IJ junctions of the bulk crystal along the *c*-axis. We attributed the smallest switching current from the supercurrent state to the resistive state to the critical current of the twisted Josephson junction, as the Josephson coupling of the artificial vdW-twisted Josephson junction should be weaker than that of the IJ junctions in the bulk. Voltage jumps appearing at higher bias current were assumed to be due to the weakened IJ coupling at the surface layers of the Bi-2212 flakes[56, 57].

To verify that the switching behavior in the *I-V* characteristics indeed originated from the Josephson coupling, we studied the Shapiro step behavior of twisted Josephson junctions. Under continuous microwave irradiation onto the device, quantized voltage steps appeared in the *I-V* characteristic curve (Figure 2c). The resonance between microwave and ac Josephson frequency resulted in these voltage steps, with a step height of *hf*/2*e*, which are called Shapiro steps[58, 59]. Shapiro steps appear as peaks in a histogram with binning of the voltage value, as shown in Figure 2c. As the microwave power increased, higher index Shapiro steps gradually appeared, as shown in Figure 2d.

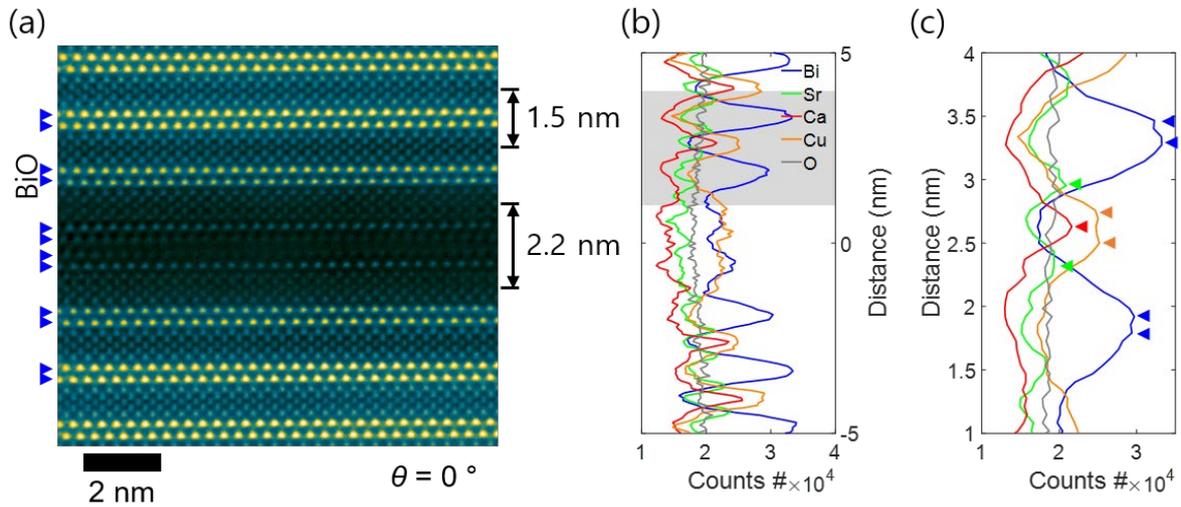

**Figure 3. Cross-sectional scanning transmission electron microscope (STEM) analysis of the twisted Josephson junction.** (a) High-angle annular dark-field (HAADF)-STEM image and (b) energy dispersive spectroscopy (EDS) result from the interface between the top and bottom Bi$_2$Sr$_2$CaCu$_2$O$_{8+x}$ flakes of the 0°-twisted Josephson junction. (c) Magnified EDS line profile of the

shaded area in (b). Atom sites appearing as peaks in the line profile are indicated by triangles in the same color as the line of the corresponding element.

To confirm the atomic interface structure, we performed scanning transmission electron microscopy (STEM) and energy dispersive spectroscopy (EDS) analyses (JEM-ARM200F; JEOL). Figure 3a shows a cross-sectional high-angle annular dark-field (HAADF)-STEM image of a twisted Josephson junction with a twist angle of 0°, the transport properties of which were measured beforehand. Due to its large atomic number (Z-contrast), the intensity of bismuth is brightest in the HAADF-STEM image. Bi-2212 crystals are known to be cleaved between BiO bilayers, so they are expected to be terminated with a BiO monolayer. However, as shown in Figure 3a, both top and bottom Bi-2212 flakes were terminated with a BiO bilayer, resulting in a BiO quadruple layer at the junction. Therefore, cleavage could occur not only at the BiO/BiO interface, but also at the SrO/BiO interface. Similar behavior has been observed in other experiments[51]. For example, if cleavage occurs at SrO/BiO, one flake would be terminated with the SrO layer, while the flake on the opposite side would be terminated with a BiO bilayer. Therefore, the artificial interface would be non-uniform over the junction area. Mixed signals of Bi and Cu at the artificial interface, as shown in Figure 3b, were attributed to this non-uniform cleavage of the Bi-2212 crystal. The periodic layered structure of the crystal is also visible in the EDS profile shown in Figure 3b. Due to the low annealing temperature of 350°C, local formation of $Bi_2Sr_2Ca_2Cu_3O_{10+x}$ near the junction interface, which often occurs during high-temperature annealing at ~865°C[60] due to intercalation of a $Ca/CuO_2$ bilayer, was not observed (Figure 3a,c). However, the center-to-center distance between $CuO_2$ bilayers of top and bottom Bi-2212 crystals in the 0°-twisted Josephson junction was about 2.2 nm, which is slightly longer than the intrinsic distance between two $CuO_2$ bilayers (~1.5 nm) in the bulk. Although no structural or chemical disorder was observed at the interface, the lower intensity in STEM and EDS images at the interface than in the bulk showed that the tunnel barrier at the interface was not uniform in the junction area.

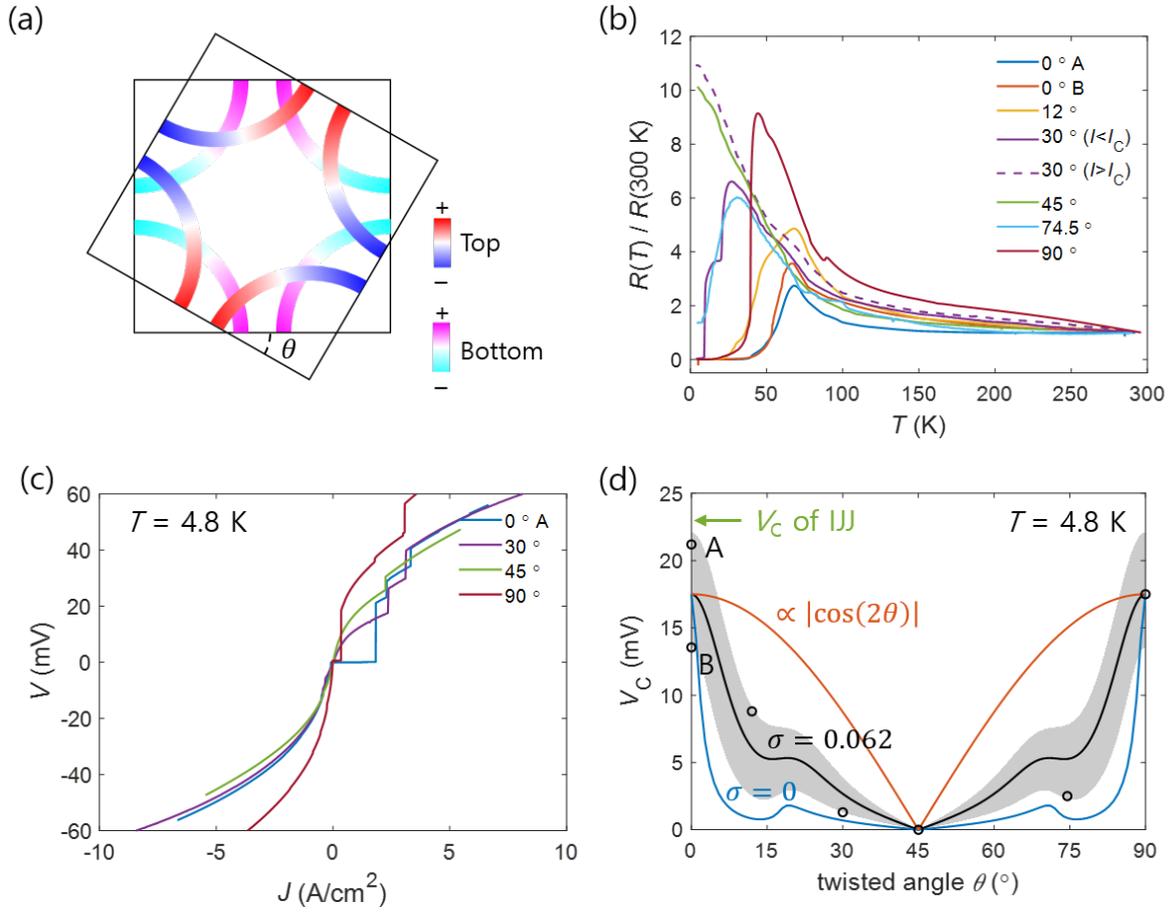

**Figure 4. Transport characteristics of twisted Josephson junctions with various twist angles.** (a) Schematic of two Fermi surfaces from the top and bottom $Bi_2Sr_2CaCu_2O_{8+x}$ (Bi-2212) crystals overlapping with a twist angle. The change from blue (cyan) to red (magenta) represents the sign and magnitude change of the order parameter in the top (bottom) Bi-2212 crystal. (b) Temperature dependence of the normalized resistance of twisted Josephson junctions with different twist angles. (c) Current-voltage (*I-V*) characteristic curve for the twisted Josephson junctions measured with sweeping bias current from negative to positive values at temperature $T = 4.8$ K. The current was normalized with the junction area. (d) Twist angle ($\theta$) dependence of the characteristic voltage ($V_c$) of twist Josephson junctions. Experimental data points are indicated as empty circles. The orange line represents the $|\cos(2\theta)|$ dependence theoretically expected when the circular Fermi surface is assumed. The black line is the best fitting of the experimental data using a tight-binding Fermi surface model and Gaussian tunneling model with tunneling coherence parameter $\sigma = 0.062$. The shaded region includes all data points with $\sigma$ from 0.037 to 0.071. The blue line represents the coherent limit ($\sigma = 0$) and the green line represents the $V_c$ jump of the IJ junction of the bulk Bi-2212.

Bi-2212 was expected to have an anisotropic *d*-wave superconducting gap, as illustrated in Figure 4a. By stacking two Bi-2212 flakes with a twist angle $\theta$ in the real space, we investigated the junction properties originating from the overlap of superconducting order parameters of each Bi-2212 flake with a relative rotation of angle $\theta$ in momentum space. This yielded valuable information about the angular dependence of the superconducting order parameter of Bi-2212 from the twist angle dependence of the Josephson coupling strength. For example, in the simplest model where the circular Fermi surface of Bi-2212 and *d*-wave anisotropy of the superconducting gap are assumed, Josephson coupling strength is expected to be proportional to $|\cos(2\theta)|$ [34].

Figure 4b shows the temperature dependence of the junction resistance normalized according to the room temperature value for comparison of different devices. All seven twisted Josephson junctions in Figure 4b show similar insulating behaviors above the superconducting transition temperature. Similar values of normalized resistance represent similar junction quality, even for different twist angles. For a device with poor junction quality, presumably due to degradation at the interface, normalized resistance increases much faster as temperature decreases, and reaches an order of magnitude larger than those in Figure 4b (see Figure S2 in the Supporting Information).

As the twist angle approaches 45°, the critical temperature of the junction decreases to almost zero due to suppression of Josephson coupling. The complete suppression of Josephson coupling of the 45°-twisted Josephson junction cannot be attributed to the poor junction quality for two reasons. First, the temperature dependence of normalized resistance for 45°-twisted Josephson junction aligns well with those of all the other junctions above the transition temperature. In addition, the normalized resistance for the 45°-twisted Josephson junction for the entire temperature range overlapped well with that of 30°-twisted junction measured with biasing current larger than the Josephson critical current $I_c$. Thus, the twist of the top and bottom crystal layers of 45° was suggested to strongly suppress Josephson coupling in the twisted junction. Second, the *I-V* characteristics of all devices, including the 45°-twisted junction, showed similar resistive branches above $I_c$ (Figure 4c). The nonlinear resistive branches in *I-V* characteristics also suggest that the twisted Josephson junction is in the tunneling regime, similar to the IJ junctions of the bulk Bi-2212.

Figure 4d shows the twist angle dependence of the characteristic voltage $V_c$, which corresponds to the junction voltage just above the critical current and represents the Josephson coupling strength. $V_c$ was used for comparison of Josephson coupling instead of $I_cR_n$ due to the uncertainty of determining normal resistance $R_n$. In another study[35], $V_c$ and $I_cR_n$ showed the same twisted angular dependence. At the 0° twist angle, the $V_c$ from two devices were similar to the value (~23 mV [56, 61]) of the IJ junctions of the bulk Bi-2212, which can be attributed to the complete overlap of superconducting order parameters in momentum space. $V_c$ decreased as the twist angle approached 45° and was fully suppressed at 45°, while $V_c$ was restored to the value of the 0° device as the twist angle approached 90°. Such twist angle dependence of Josephson coupling is consistent with the anisotropic *d*-wave superconducting order in Bi-2212.

In the simple model assuming a circular Fermi surface of Bi-2212, the *d*-wave superconducting order should give an angular dependence of $V_c \sim |\cos(2\theta)|$ [34]. However, the concave upward shape of the observed twist angle dependence of $V_c$ was in contrast to the concave downward angular dependence of $|\cos(2\theta)|$. That is, $V_c$ decreased as the twist angle approached 45° faster than the case of the circular Fermi surface model. To quantitatively explain the observed angular dependence of $V_c$, we adopted the tight-binding Fermi surface model, which is more realistic than the circular Fermi surface model[34, 51, 62, 63]. Josephson coupling occurs by the overlap between the top and bottom Bi-2212 Fermi surfaces. As depicted in Figure 4a, the tight-binding Fermi surface is not rotationally symmetrical. Therefore, angle twisting between the top and bottom layers reduces the overlap of the Fermi surfaces more significantly compared to the case of the circular Fermi surface with rotational symmetry. This qualitatively explained the observed twist angle dependence. In addition, we introduced finite tunneling incoherence to consider the nonuniformity of the tunnel barrier at the junction interface shown in STEM images (Figure 3).

The dependence of the critical current on the twist angle $\theta$ can be theoretically obtained by computing the critical current of a weakly coupled Josephson junction[34]:

$$J_c(\theta) = \left| 4eT \sum_\omega \int_{FBZ} d^2k d^2k' f(\mathbf{k}, \mathbf{k}') F_\omega(\mathbf{k}) F_\omega^\dagger(\tilde{\mathbf{k}}') \right|$$

where $e$ is the electronic charge, $T$ is the temperature, and $\omega$ is the fermionic Matsubara frequency. FBZ is the first Brillouin zone, which is parameterized by the momentum vector $\mathbf{k} = (k_x, k_y)$ perpendicular to the *c*-axis. $f(\mathbf{k}, \mathbf{k}')$ is the tunneling matrix element between the flakes, $\tilde{\mathbf{k}}'$ is $\mathbf{k}'$ rotated by the twist angle $\theta$ about the *c*-axis. $F_\omega(\mathbf{k})$ is the anomalous Green's function. The details of the computation of $J_c(\theta)$, including definitions of $f(\mathbf{k}, \mathbf{k}')$ and $F_\omega(\mathbf{k})$, are presented in the Supporting Information. The tunneling incoherence is included phenomenologically by $f(\mathbf{k}, \mathbf{k}') = f_0 \exp[-|\mathbf{k} - \mathbf{k}'|^2 a^2 / 2\pi^2 \sigma^2]$, where incoherence is quantified by the tunneling incoherence parameter σ [34]. The tunneling incoherence parameter σ of 0 (∞) represents the fully coherent (incoherent) limit. The calculated curve best fits the experimental data with σ = 0.062 (black line in Figure 4d), and all the data points are in the range of σ from 0.037 to 0.071 (shaded area in Figure 4d).

In summary, we investigated the anisotropic superconducting order parameter of Bi-2212 by studying the twist angle dependence of the Josephson coupling of twisted Josephson junctions. We adopted the microcleave-and-stack technique so that the surfaces of the Bi-2212 flakes forming the twisted Josephson junctions did not come into contact with any other materials. The temperature dependence of the resistance and current-voltage characteristics showed the tunneling behavior of twisted Josephson junctions. Josephson coupling was further confirmed by observing Shapiro steps under microwave irradiation. The Josephson coupling was comparable to that of IJ junctions of bulk Bi-2212 for twist angles of 0° and 90°, but fully suppressed for 45°. The observed twist angle dependence was quantitatively explained by the model calculation considering the *d*-wave superconducting order, tight-binding Fermi surface of Bi-2212, and finite tunneling incoherence at the junction, consistent with the nonuniformity of the tunnel barrier at the junction interface, as observed by STEM and EDS. Our technique also provides the possibility of studying exotic topological superconductivity in the twist angle near 45° [64]. We demonstrated that the microcleave-and-stack technique is a promising fabrication method compatible with sensitive layered materials, widening the applications of the vdW-based twistronics platform.

**AUTHOR INFORMATION**

Corresponding Author

*lghman@postech.ac.kr (G.-H.L.)

*hjlee@postech.ac.kr (H.-J.L.)

Author Contributions

G.-H.L. and H.-J.L. conceived and supervised the project. J.L. fabricated the samples and performed transport measurement with help of Y.-B.C., J.P., and J.S.. G.G. provided the Bi-2212 crystals. W.L. and G.Y.C. performed theoretical analysis and calculations. G.-Y.K. and S.-Y.C. performed STEM and EDS analysis. J.L., W.L., G.-Y.K., S.-Y.C., G.Y.C., G.-H.L., and H.-J.L. wrote the paper.

Funding Sources

J.L., J.P. and H.-J.L. were supported by the National Research Foundation (NRF) through the SRC Center for Topological Matter, POSTECH, Korea (Grant No. 2018R1A5A6075964). Y.-B.C., S.J. and G.-H.L. acknowledge the support of National Research Foundation (NRF) funded by the Korean

Government (Grant No. 2020R1C1C1013241 and 2020M3H3A1100839), Samsung Science and Technology Foundation (Project No. SSTF-BA1702-05), Samsung Electronics Co., Ltd (IO201207-07801-01) and Basic Science Research Institute Fund (Grant No. 2021R1A6A1A10042944). W.L. and G.Y.C. acknowledge the support of the National Research Foundation of Korea (NRF) funded by the Korean Government (Grant No. 2020R1C1C1006048 and 2020R1A4A3079707), as well as Grant No. IBS-R014-D1. G.Y.C. is also supported by the Air Force Office of Scientific Research under Award No. FA2386-20-1-4029. The work at BNL was supported by the US Department of Energy, office of Basic Energy Sciences, contract no. DOE-sc0012704. G.-Y.K. and S.-Y.C. acknowledges the support of the Global Frontier Hybrid Interface Materials of the National Research Foundation of Korea (NRF) funded by the Ministry of Science and ICT (2013M3A6B1078872), and Korea Basic Science Institute (National research Facilities and Equipment Center) Grant (2020R1A6C101A202) funded by the Ministry of Education.

Notes

The authors declare no competing financial interest.

# Supporting Information
# Twisted van der Waals Josephson junction based on high-$T_c$ superconductor

Jongyun Lee[1], Wonjun Lee[1], Gi-Yeop Kim[2,3], Yong-Bin Choi[1], Jinho Park[1], Seong Jang[1], Genda Gu[4], Si-Young Choi[2,3], Gil Young Cho[1,5,6], and Gil-Ho Lee[1,6,*], Hu-Jong Lee[1,*]

[1] Department of Physics, Pohang University of Science and Technology, Pohang 37673, Korea

[2] Department of Materials Science and Engineering, Pohang University of Science and Technology, Pohang 37673, Korea.

[3] Materials Imaging & Analysis Center, Pohang University of Science and Technology, Pohang 37673 Korea

[4] Condensed Matter Physics and Materials Science Department, Brookhaven National Laboratory, Upton, New York 11973, USA

[5] Center for Artificial Low Dimensional Electronic Systems, Institute for Basic Science (IBS), Pohang 37673, Korea

[6] Asia Pacific Center for Theoretical Physics, Pohang 37673, Korea

# 1. Characteristics of all twisted Bi$_2$Sr$_2$CaCu$_2$O$_{8+x}$ (Bi-2212) van der Waals (vdW) Josephson junctions

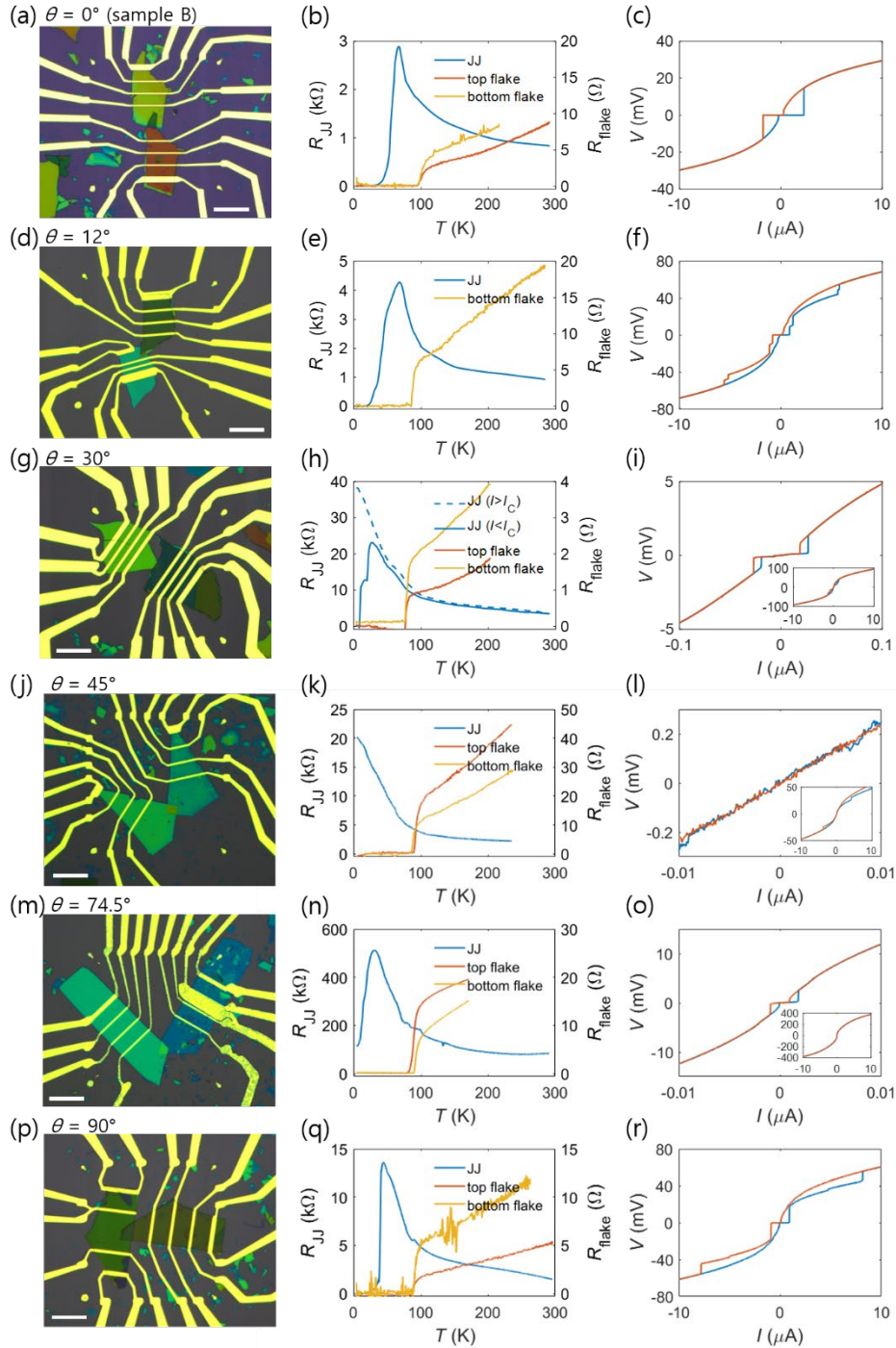

**Figure S5. Basic transport characteristics of all twisted Josephson junctions.** (a), (d), (g), (j), (m), (p) are the optical images with scale bars of 50 μm, (b), (e), (h), (k), (n), (q) are the temperature dependent resistances, and (c), (f), (i), (l), (o), (r) are the current-voltage (*I-V*) curves at 4.8 K of the Bi$_2$Sr$_2$CaCu$_2$O$_{8+x}$ (Bi-2212) vdW Josephson junctions with the twist angle of 0° (sample B), 12°, 30°, 45°, 74.5° and 90°, respectively.

## 2. Characteristics of twisted junctions with poor interface quality

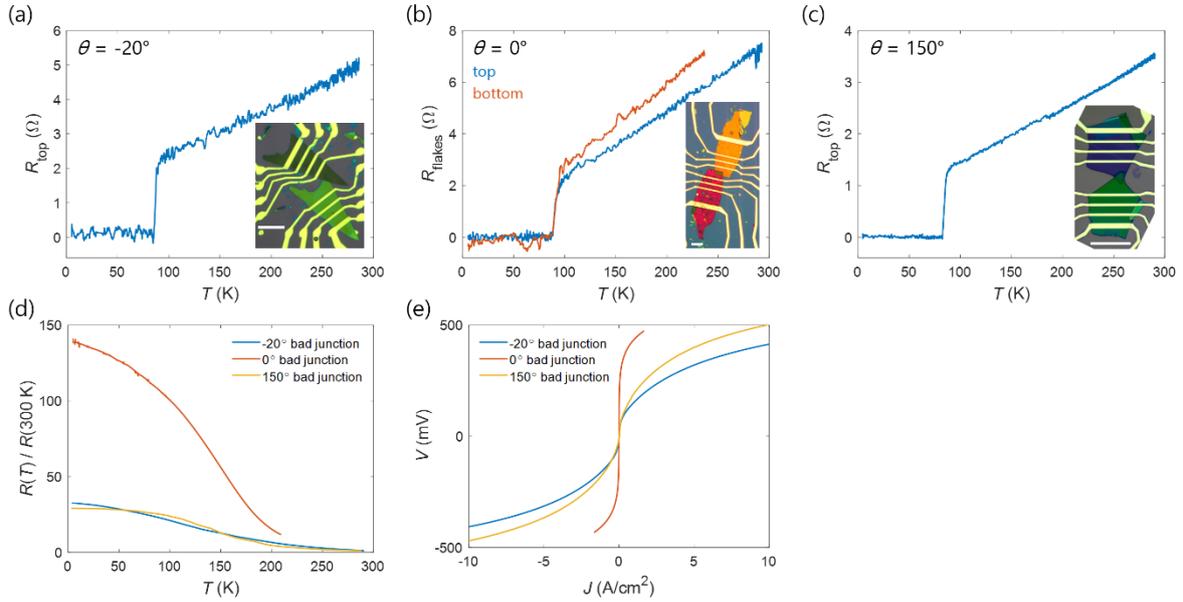

**Figure S2. Transport characteristics of twisted junctions with poor interface quality.** (a), (b), (c) The temperature-dependent resistance of the Bi-2212 flakes consisting the twisted junctions with angle of -20°, 0° and 150°, respectively. The inset of each figure is an optical image of the corresponding sample, with scale bar of 50 μm. (d) Temperature dependence of normalized resistance of twisted junctions with poor junction interfaces. (e) $I$-$V$ characteristic curve for the twisted junctions with poor junction interfaces. Current was normalized with the junction area and bias current was swept from negative to positive values at the temperature $T$ = 4.8 K.

As described in the main text, junctions for which lock-in measurements were not possible due to poor junction quality were excluded from the angle dependence comparison. Therefore, we measured temperature-dependent resistance of those poor-quality junctions with dc measurement, instead of lock-in measurement. On the other hand, the contact resistance of Ag/Au electrodes was small enough for lock-in measurement. Therefore, it was possible to measure the resistance of each flake by lock-in measurement, with the current path not passing across the twisted junction.

## 3. Details of the theoretical calculation

The twist angle dependence of the critical current density for a stack of weakly coupled Josephson junctions is described by[1]

$$J_c(\theta) = \left| 4eT \sum_{\omega} \int_{FBZ} d^2k\, d^2k'\, f(\mathbf{k}, \mathbf{k'}) F_\omega(\mathbf{k}) F_\omega^\dagger(\mathbf{\tilde{k}'}) \right|$$

where $e$ is the electronic charge, $T$ is the temperature, $\omega$ is the fermionic Matsubara frequency, 'FBZ' stands for the first Brillouin zone, $\mathbf{k} = (k_x, k_y)$ is the momentum vector perpendicular to the $c$-axis, $f(\mathbf{k}, \mathbf{k'})$ is the tunneling matrix element, $\mathbf{\tilde{k}'}$ is $\mathbf{k'}$ rotated by the given angle $\theta$ about the $c$-axis, and $F_\omega(\mathbf{k})$ is the anomalous Green's function.

Let us first explain the detailed functional forms of $f(\mathbf{k}, \mathbf{k'})$ and $F_\omega(\mathbf{k})$[1]. Here, the tunneling matrix element is given by a finite-width Gaussian function

$$f(\mathbf{k}, \mathbf{k'}) = f_0 \exp\left[ -\frac{|\mathbf{k} - \mathbf{k'}|^2 a^2}{2\pi^2 \sigma^2} \right],$$

where $a$ is the lattice spacing, $f_0$ is the tunneling amplitude, and $\sigma$ parametrizes the incoherence. This tunneling matrix effectively allows a process violating momentum conservation. It turns out that the root mean square of the error, the difference between the theoretical prediction of $J_c(\theta)/J_c(0)$ and our experimental data, is minimized with σ = 0.062 and 4$Tf_0$=17.4 meV, and that all the data points are in the region bounded by the two curves, one with (σ=0.037, 4$Tf_0$=13.55 meV) and the other with (σ=0.071, 4$Tf_0$=21.2 meV), respectively. On the other hand, the anomalous Green's function is given by the standard BCS theory

$$F_\omega(\mathbf{k}) = \frac{\Delta_\mathbf{k}(T)}{[\omega^2 + \xi_\mathbf{k}^2 + |\Delta_\mathbf{k}(T)|^2]},$$

with the superconducting order parameter $\Delta_\mathbf{k}(T)$ and the quasiparticle dispersion $\xi_\mathbf{k}$. Here, the superconducting order parameter is assumed to be the conventional $d$-wave one,

$$\Delta_\mathbf{k}(T) = \Delta_0 [\cos(k_x a) - \cos(k_y a)],$$

where the temperature dependent gap $\Delta_0(T)$ is approximated as $\Delta_0 \approx \Delta_0(0) = 5.1$ meV since the temperature $T = 4.8$ K of the sample is much smaller than $T_c = 85.8$ K, i.e. $T \ll T_c$. The quasiparticle dispersion is taken from the tight-binding model[1],

$$\xi_\mathbf{k} = -t[\cos(k_x a) + \cos(k_y a)] + t' \cos(k_x a) \cos(k_y a) - \mu$$

with $t = 306$ meV, $t'/t = 0.9$, and $\mu/t = -0.675$.

The computation of $J_c(\theta)$ is conducted by first discretizing the FBZ into $N$-by-$N$ cells with $N = 500$, and summing up the integrands over all the cells. After the integration, the Matsubara frequency summation is taken for the first $N_\omega = 600$ fermionic frequencies, $\omega_n = (2n + 1)\pi/\beta$ with the inverse temperature $\beta = \hbar/k_\text{B}T$.